\documentclass[twocolumn,pra,amssymb,showpacs]{revtex4-1}
\usepackage{graphicx}
\usepackage{amsmath}
\usepackage{subfigure}

\newcommand{\be}{\begin{equation}}
\newcommand{\ee}{\end{equation}}
\newcommand{\ben}{\begin{eqnarray}}
\newcommand{\een}{\end{eqnarray}}

\newcommand{\nd}{\noindent}

\begin{document}

\title{New approach to finding the maximum number of mutually unbiased bases in $\mathbb{C}^6$}
\author{J. Batle$^{1}$}
\email{E-mail address: jbv276@uib.es}
\affiliation{ $^1$Departament de F\'{\i}sica, Universitat de les Illes Balears,
 07122 Palma de Mallorca, Balearic Islands, Europe  \\\\}
 
\date{\today}

\begin{abstract}

\nd There has been great interest in finding sets of $m$ mutually unbiased bases which are compatible with a 
given space $\mathbb{C}^d$, specially in physics due to their interesting applications in quantum information theory. 
Several general results have been obtained so far, but surprising results may occur for definite $(m,d)$-values. 
One such case that has remained an open question (the simplest case) is the one regarding the existence of $m=4$ mutually orthogonal 
bases for $d=6$. In the present work we introduce a new approach to the problem by translating it into an optimization 
procedure for a given pair $(m,d)$.
\end{abstract}

\pacs{03.65.-w,03.67.-a,03.65.Ta} 

\maketitle

\section{Introduction}

The paradigm of observables defined on an infinite Hilbert space being mutually incompatible 
in quantum mechanics is provided by the Heisenberg commutation relations for the position and momentum 
operators. The associated Heisenberg group --in connection with the corresponding 
Weyl algebra-- of phase-space 
translations is still relevant for systems with a finite number of
orthogonal states, providing a basis of the space $\mathbb{C}^d$. 
As first studied by Schwinger, for each
dimension $d \geq 2$ there is a set of unitary operators which give rise to
a discrete equivalent of the Heisenberg-Weyl group \cite{schwinger60}. 

We may somehow a priori expect that the kinematics of composite quantum systems will not 
depend on the dimensions of their building blocks. That is, that composite systems let us say consisting 
in the tensor product of two different Hilbert spaces (differing only in the corresponding dimensions) 
will undergo a similar evolution since they are structurally identical. Mathematically, the previous fact 
would imply the state spaces $\mathbb{C}^d$ to be
structurally identical, at least with respect to properties closely related
to the Heisenberg-Weyl group. 

However, it is surprising that the aforementioned group allows one to construct 
$(d+1)$ so-called mutually unbiased (MU) bases of the space $\mathbb{C}^d$ if $d$ is the power of a
prime number \cite{ivanovic81,wootters+89}, whereas the construction fails in
all other dimensions. In point of fact, no other successful method to construct $(d+1)$ MU bases in 
all dimensions is known \cite{archer05,planat+06}.

Given $m = d+1$ orthonormal bases in the space $\mathbb{C}^{d}$, they are 
\textit{mutually unbiased} if the moduli of the scalar products among the $d(d+1)$ basis vectors take these values: 

\begin{equation}  \label{MUB}
\left\vert {\langle \psi _{j}^{b}|}{\psi _{j^{\prime }}^{b^{\prime }}
\rangle }\right\vert =\left\{ 
\begin{array}{ll}
\delta_{jj^{\prime }} & \quad \mbox{if $b = b^\prime$}\, , \\ 
\frac{1}{\sqrt{d}} & \quad \mbox{if $b \neq b^\prime $}\,,%
\end{array}
\right. 
\end{equation}

\noindent where $b,b^{\prime }=0,1,\ldots ,d$. MU bases have useful applications in 
many quantum information processing. Such (complete) sets of MU bases
are ideally suited to reconstruct quantum states \cite{wootters+89} while
sets of up to $(d+1)$ MU bases have applications in quantum cryptography 
\cite{cerf+02, brierley09} and in the solution of the mean king's problem \cite{aharonov+01}. 
Even for $d=6$, we do not know whether there exist four MU bases
or not \cite{s6,s7,s8,s9}. Hence the research on the maximum number of bases for $d=6$ and construction 
of MU bases in $\mathbb{C}^{6}$ is
of great importance. The issue of MU bases constitutes another part in the field of quantum information theory 
that is involved in pure mathematics, such as number theory, abstract algebra and projective 
algebra.

The methods to construct complete sets of MU bases typically deal with all
prime or prime-power dimensions. They are constructive methods and effectively 
lead to the same bases. Two (or more) MU bases thus correspond to two (or more) unitary
matrices, one of which can always be mapped to the identity $I$ of the space 
$\mathbb{C}^{d}$, using an overall unitary transformation. It then follows from the
conditions (\ref{MUB}) that the remaining unitary matrices must
be complex Hadamard matrices: the moduli of all their matrix elements equal 
$1/\sqrt{d}$. This representation of MU bases links their classification to
the classification of complex Hadamard matrices \cite{complex}.

In this paper, we choose a different method to study MU bases in
dimension six or any other dimension $d$. We will approach the problem by directly exploring the unitary matrices 
-- randomly distributed, but according to the Haar measure -- 
whose columns vectors constitute the bases elements, which must fulfill a series 
of requirements concerning their concomitant bases being unbiased. The overall scenario 
reduces to a simple --though a bit involved-- optimization procedure. In point of fact, we shall perform a 
two-fold search employing i) an amoeba optimization procedure, where the optimal value 
is obtained at the risk of falling into a local minimum and ii) the so called simulated annealing \cite{kirkpatrick83} 
well-known search method, a Monte Carlo method, inspired by the cooling processes of molten metals. 
The advantage of this duplicity of computations is that we can be absolutely confident about the final 
result reached. Indeed, the second recipe contains a mechanism 
that allows a local search that eventually can escape from local optima.

This paper is organized as follows. In Section II we describe the generation of unitary matrices 
according to their natural Haar measure. Section III explains how the optimization is performed and 
the concomitant results are shown in Section IV. Finally, some conclusions are drawn in Section V.

\section{The Haar measure and the concomitant generation of ensembles of random matrices} 

The applications that have appeared so far in quantum information theory, 
in the form of dense coding, teleportation, quantum cryptography and specially 
in algorithms for quantum computing (quantum error correction codes for instance), 
deal with finite numbers of qubits. A quantum gate which acts upon these qubits 
or even the evolution of 
that system is represented by a unitary matrix $U(N)$, with $N=2^n$ being the 
dimension of the associated Hilbert space ${\cal H}_N$. The state $\rho$ describing 
a system of $n$ qubits is given by a hermitian, positive-semidefinite ($N \times N$) 
matrix, with unit trace. In view of these facts, it is natural 
to think that an interest has appeared in the {\it quantification} of 
certain properties of these systems, most of the 
times in the form of the characterization of a certain state $\rho$, 
described by $N \times N$ matrices of finite size. Natural applications arise 
when one tries to simulate certain processes through random matrices, 
whose probability distribution ought to be described accordingly.

This enterprise requires a quantitative measure $\mu$ on a given set of 
matrices. There is one natural candidate measure, 
the {\bf Haar measure} on the group ${\cal U}(N)$ of unitary matrices. 
In mathematical analysis, the Haar measure \cite{Haar33} 
is known to assign an ``invariant volume" to what is known as subsets of locally compact topological groups. 
Here we present the formal 
definition \cite{Conway90}: given a locally compact topological group 
$G$ (multiplication is the group operation), consider a $\sigma$-algebra $Y$ 
generated by all compact subsets of $G$. 
If $a$ is an element of $G$ and $S$ is a set in $Y$, then the set 
$aS = $ $\{$ $as : s \in S$ $\}$ also belongs to $Y$. A measure $\mu$ on $Y$ will be 
letf-invariant if $\mu(aS)=\mu(S)$ for all $a$ and $S$. Such an invariant measure 
is the Haar measure $\mu$ on $G$ (it happens to be both left and 
right invariant). In other words \cite{Haarsimetria}, the Haar measure defines 
the unique invariant integration measure for Lie groups. It implies that a 
volume element d$\mu(g)$ is identified by defining the integral of a function 
$f$ over $G$ as $\int_G f(g) d\mu(g)$, being left and right invariant 

\begin{equation}
\int_G f(g^{-1}x) d\mu(x)\,=\,\int_G f(x g^{-1}) d\mu(x)\,=\,\int_G f(x) d\mu(x).
\end{equation}

\noindent The invariance of the integral follows from the concomitant invariance 
of the volume element d$\mu(g)$. It is plain, then, that once d$\mu(g)$ is fixed 
at a given point, say the unit element $g=e$, we can move performing a 
left or right translation. 

We do not gain much physical insight with these definitions of the Haar measure and its 
invariance, unless we identify 
$G$ with the group of unitary matrices ${\cal U}(N)$, the element $a$ with a 
unitary matrix $U$ and $S$ with subsets of the group of unitary matrices ${\cal U}(N)$, 
so that given a reference state $|\Psi_0\rangle$ and a unitary matrix $U \in {\cal U}(N)$, 
we can associate a state $|\Psi\rangle_0=U|\Psi_0\rangle$ to $|\Psi_0\rangle$.
Physically what is required is a probability measure $\mu$ invariant under 
unitary changes of basis in the space of pure states, that is, 

\begin{equation}
P^{(N)}_{Haar}(U\,|\Psi\rangle)\,=\,P^{(N)}_{Haar}(|\Psi\rangle).
\end{equation}

\noindent These requirements can only be met by the Haar measure, which is 
rotationally invariant.
\newline

Now that we have justified what measure we need, we should be able to generate  
random matrices according to such a measure in arbitrary dimensions. 
The theory of random matrices \cite{Mehta90} specifies different 
{\it ensembles} of matrices, classified according to their different 
properties. In particular, the Circular Unitary Ensemble (CUE) consists 
of all matrices with the (normalized) Haar measure on the unitary 
group ${\cal U}(N)$. The Circular Orthogonal Ensemble (COE) is described 
in similar terms using orthogonal matrices, and it was useful in order 
to describe the entanglement features of two-{\it rebits} systems. Given 
a $N \times N$ unitary matrix $U$, the minimum number of independent entries 
is $N^2$. This number should match those elements that need to describe 
the Haar measure on ${\cal U}(N)$. This is best seen from the following 
reasoning. Suppose that a matrix $U$ is decomposed as a product of two 
(also unitary) matrices $U = X Y$. In the vicinity of $Y$, we have 
\cite{Mehta90} $U + dU = X(1 + idK)Y$, where $dK$ is a hermitian matrix 
with elements $dK_{ij} = dK^{R}_{ij} + idK^{I}_{ij}$. Then the probability 
measure nearby $dU$ is $P(dU) \sim \prod_{i \le j}dK^{R}_{ij} 
\prod_{i < j}dK^{I}_{ij}$, which accounts for the number of independent 
variables. Such measure for CUE is invariant \cite{Mehta90} and therefore 
proportional to the Haar measure. 

Yet, the aforementioned description is not useful for practical purposes. We 
need to parameterize the unitary matrices according to the Haar measure. 
According to the parameterization for CUE dating back to Hurwitz \cite{Hur1887} using Euler angles, 
the basic assumption is that an arbitrary unitary matrix can be decomposed into 
elementary two-dimensional transformations, denoted by 
$E^{i,j}(\phi,\psi,\chi)$: 

\begin{eqnarray} \label{Eij} 
E^{i,j}_{kk} &=& 1 \,\,\,\,\,\,\,\,\,\,\,\,\,\,\,\,\,\,\,\, 
k=1, .., N; \,\,\,\,\,\,\,\,\,\, k \neq i,j \cr
E^{i,j}_{ii} &=& \cos \phi \, e^{i\psi} \cr
E^{i,j}_{ij} &=& \sin \phi \, e^{i\chi} \cr
E^{i,j}_{ji} &=& -\sin \phi \, e^{-i\chi} \cr
E^{i,j}_{jj} &=& \cos \phi \, e^{-i\psi}. 
\end{eqnarray} 

\noindent Using these elementary rotations we define the composite 
transformations

\begin{eqnarray} \label{Es}
E_{1} &=& E^{N-1,N}(\phi_{01},\psi_{01},\chi_{1}) \cr
E_{2} &=& E^{N-2,N-1}(\phi_{12},\psi_{12},0) 
E^{N-1,N}(\phi_{02},\psi_{02},\chi_{2}) \cr
E_{3} &=& E^{N-3,N-2}(\phi_{23},\psi_{23},0) 
E^{N-2,N-1}(\phi_{13},\psi_{13},0) \cr 
 && E^{N-1,N}(\phi_{03},\psi_{03},\chi_{3}) \cr
... &=& ... \cr
E_{N-1} &=& E^{1,2}(\phi_{N-2,N-1},\psi_{N-2,N-1},0) \cr 
&& E^{2,3}(\phi_{N-3,N-1},\psi_{N-3,N-1},0)... \cr
&& E^{N-1,N}(\phi_{0,N-1},\psi_{0,N-1},\chi_{N-1}),
\end{eqnarray} 

\noindent we finally form the matrix 

\begin{equation} \label{U}
U \,=\, e^{i\alpha}\,E_1 E_2 E_3 ... E_{N-1}
\end{equation}

\noindent with the angles parameterizing the rotations 

\begin{equation} \label{angles}
0 \le \phi_{rs} \le \frac{\pi}{2} \,\,\,\, 0 \le \psi_{rs} < 2\pi 
\,\,\,\, 0 \le \chi_{1s} < 2\pi \,\,\,\, 0 \le \alpha < 2\pi.
\end{equation}
  
\noindent The ensuing (normalized) Haar measure \cite{Girko90}

\begin{eqnarray} \label{PHaar}
P_{Haar}(dU)\,&=&\,\sqrt{N!2^{N(N-1)}}d\alpha \cr 
&& \prod_{1 \le r < s \le N} \frac{1}{2r} d [(\sin \phi_{rs})^{2r}] d\psi_{rs} \cr
&& \prod_{1 < s \le N} d\chi_{1s}
\end{eqnarray}

\noindent provides us with a random matrix belonging to CUE. 
Finally, we randomly generate the angles (\ref{angles}) {\it uniformly} 
and obtain the desired random matrix $U$ (\ref{U}).

\section{Description of the optimization procedure} 

Let us formulate the problem of having $m$ orthonormal bases $B_i,i=1..m$ in terms of the 
elements of a unitary matrix. All basis elements or vectors are obtained from a unitary matrix by identifying them 
with the corresponding columns. Unitarity guarantees that all vectors will therefore be orthonormal. 
Now we have to cope with the bases being unbiased amidst them. Since each basis is represented by a unitary 
matrix, we then have $B_i,i=1..m\, \rightarrow \, U_i,i=1..m$. This condition can be addressed by imposing that 
matrix elements 

\begin{equation} \label{Ucdot}
\big(U_i \cdot U_j \big)_{lm},
\end{equation}

\noindent where $i=1 < j \leq m$, have to be equal to $1/\sqrt{d}$. In other words, 
$U_i \cdot U_j$ has to be proportional to a Hadamard-like matrix. The aforementioned 
conditions has to be applied to all possible $m(m-1)/2$ bipartite combinations of bases $B_i$. 

Let us define the following quantities as the {\it residuals} 

\begin{equation}
\rho_{l,m,i,j} \equiv  \bigg(\left\vert \big(U_i \cdot U_j \big)_{l,m} \right\vert^2 \,-\, 1/d \bigg)^2.
\end{equation}

\noindent Thus, the problem of finding a set of unbiased orthonormal bases is translated into the optimization 
procedure of finding the minimum of $\sum_{l,m,i,j}\,\rho_{l,m,i,j}$ being equal to zero. If the minimum is different from 
zero, given $d$ and $m$, we definitely do not have a set of unbiased bases. In addition, our function resembles 
very much the quantity used in \cite{metrics} to define the notion of ``unbiasedness'' between two orthonormal bases. 
To whether or not the aforementioned quantity represents a metric is something not checked.

Now that the we have translated the problem of finding MU bases into an operational one, one has to be able to 
explore all possible bases. This fact means that we have to be able to survey the set space of unitary matrices. 
Since we described in the previous section how to generate random unitary matrices properly, we will have to 
numerically explore all unitary matrices. The way to pursue that is to consider the angles (\ref{angles}) --given $d$ and $m$-- 
in all cases in (\ref{Ucdot}) as the variables of the function $\sum_{l,m,i,j}\,\rho_{l,m,i,j}$ to be minimized. Provided the 
concomitant optimal value (the sum of all residuals $\rho_{l,m,i,j}$) is equal to zero, we may then have found a set of MU bases. 
Otherwise, that may not be possible given the constraints on $d$ and $m$.

\section{Results} 

Now that we have the tools to perform a numerical survey over the set of unitary matrices, we carry out the optimization 
described in the previous section. 

\subsection{d=6, m=3} 

\begin{figure}[htbp]
\begin{center}
\includegraphics[width=8.8cm]{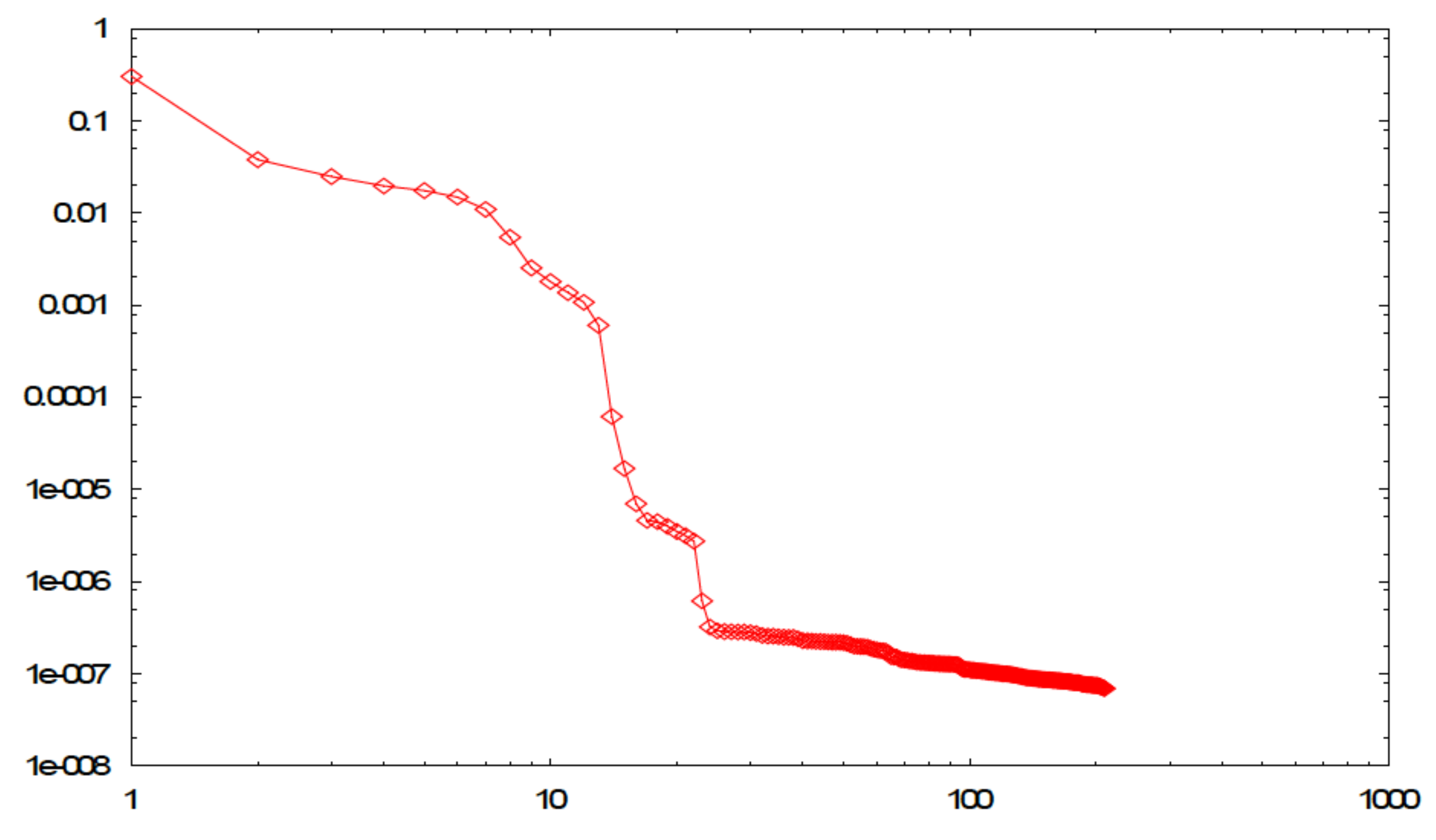}
\caption{(Color online) Plot of the evolution of the sum of residuals --the figure of merit in the optimization-- for the $d=6$ case with three basis vs. the number of 
Monte Carlo steps. As can be clearly appreciated, a global zero minimum is reached. See text for details.}
\label{FigA}
\end{center}
\end{figure}

The $\mathbb{C}^6$ case with three bases is know to exist, so our numerical procedure must return a minimal value of zero. 
The results are depicted in Fig. (\ref{FigA}). As can be appreciated, convergence is reached very fast after each Monte Carlo step 
(formed by 15000 different configurations each). Therefore, we are quite confident that we have found a set of MU bases in the 
$(d=6,m=3)-$case. 

However, we must bear in mind an important issue regarding numerical surveys. Not all our simulations lead to a zero minimum, so the lack of convergence 
is in favor of the argument that some sets of MU bases cannot be extended to further number of bases. 
In $d=6$ there are some sets of 2 MU bases that cannot be extended to 3 MU bases (see Ref. \cite{Dardo} and references therein). 
From numerical simulations it is knows that null measure sets cannot be reached. In \cite{Dardo}, 
a subset of the Karlson's family of complex Hadamard matrices
cannot be extended to 3 MU bases. Additionally, the Karlson's family has
dimension 2 and the maximal set of complex Hadamard matrices in dimension
6 has dimension 4, so it is a null measure set. Therefore, one could never achieve a unitary matrix from 
random simulations such that it belongs to the Karlsson's family. Moreover, there are 1
dimensional families and even more, isolated complex Hadamard matrices in dimension six.

When considering the extension of the number of MU bases $\{I,H_i\}$, $H_i$ being a Hadamard matrix, 
provided by a certain number, we then know that that function lacks the property of continuity \cite{Jaming}. 
The absence of continuity together with an 
incomplete knowledge about the number discontinuities in the number of MU bases makes the overall 
problem a difficult one. However, in our approach, we succeed in finding at least a few cases where 
$(m=3,d=6)$ holds.

The study of the case with three bases confirms that our approach to the problem is a good one. As a matter of fact, we could study the problem for 
any $(d,m)-$case, but the overall optimization procedure --as it is indeed the case for any simulation of a quantum system-- becomes intractable 
at some point. With the numerical tools being a valid one, we can now tackle the problem of whether $\mathbb{C}^6$ can sustain $m=4$ MU bases.

\subsection{d=6, m=4}

\begin{figure}[htbp]
\begin{center}
\includegraphics[width=8.8cm]{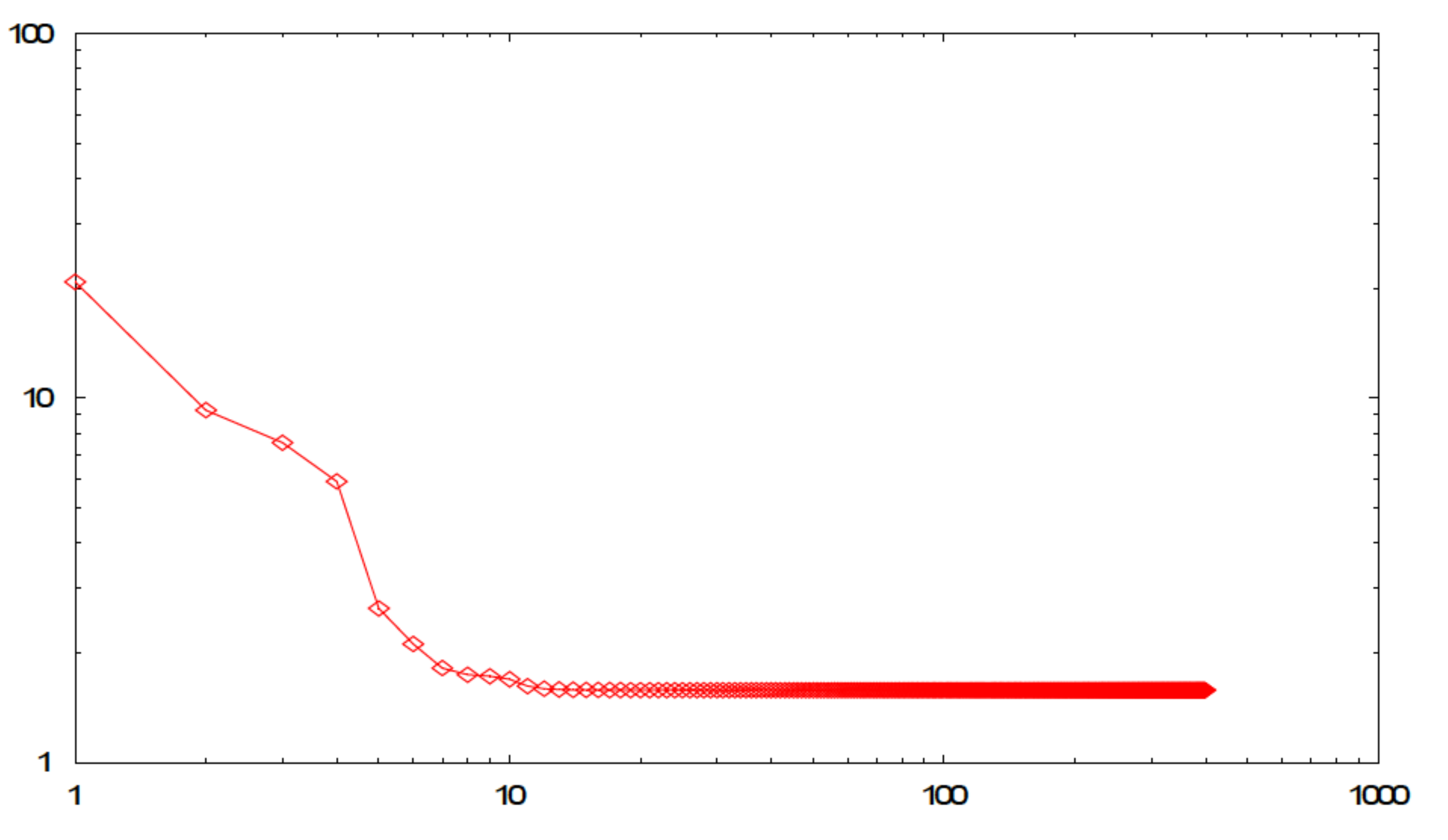}
\caption{(Color online) Plot of the evolution of the sum of residuals for the $d=6$ case with four basis vs. the number of 
Monte Carlo steps. This typical evolution of the function to be optimized --the sum of residuals in our case-- 
does not reach a minimum of zero. It approaches zero but the corresponding value is aways of $O(1)$. See text for details.}
\label{FigB}
\end{center}
\end{figure}

Now that we have implemented the tools for performing a search in the space of unitary matrices of a given dimension $N \times N$, we are 
in a position a bit closer to ascertain whether it is possible to have four MU bases in the $d=6-$case. We start the numerical search and the outcome of if is shown 
in Fig. (\ref{FigB}). The evolution is such that the total function to be minimized rapidly decreases, and attains a value that is not zero. Several repetitions of the 
same optimization procedure lead to the same conclusion: the value which is optimized is of $O(1)$. Thus, we have more evidence that 
four mutually unbiased bases cannot occur in $\mathbb{C}^6$. However, in the light of the previous discussion on continuity, it still remains 
doubts that our numerical procedure may not arrive at the minimum of 0 because we are trying to explore a set of zero measure. This fact 
implies that our numerical approach to the problem may have (still) some loopholes as far as reaching a conclusive answer. 
All facts points towards that $m=4$ is incompatible with $d=6$, but we have no theorem that ascertains whether the function 
which is optimized reaches may ever reach a minimum of zero.

In addition, we are left with an intriguing question: what is the meaning of having a set of four almost MU bases? 
(let us call them $\epsilon-$MU bases from now on). Definitely, if we have found one such $\epsilon-$MU bases set, it may not be unique. In point of fact, 
there may exist as many as different vales for the function to be optimized are reached. However, what is the physics that entails that one family of these 
$\epsilon-$MU bases reaches a minimum minimorum? In operational terms, what role could these $\epsilon-$MU bases play in practice? It may be the 
case, for instance, that a subset of the four bases is mutually unbiased.

\section{Conclusions}

We have translated the problem of the existence of $m=4-$bases in $\mathbb{C}^6$ into an optimization procedure. As expected, 
the concomitant numerical optimization has provided a satisfactory answer for known cases such as $m=3$ in $\mathbb{C}^6$. 
This new approach to the problem of to whether or not there exist a set of $m=4$ MU bases for $d=6$ has provided more evidence 
in favor that this is not case, although no theorem guarantees this argument. 

In addition, we are left with the interesting question on the limitations that pose the use of sets of imperfect MU bases in 
quantum information tasks, an issue that is certainly of interest for in experiments one has to deal with imperfections. 
Also, our procedure is capable to explore more dimensions and bases in a straightforward manner, although taking 
into account that a computational limitation is reached, and therefore opens the door to similar studies in the future \cite{future}.

\section*{Acknowledgements}

J. Batle acknowledges partial support from the Physics Department, UIB. J. Batle acknowledges fruitful discussions with D. Goyeneche, 
J. Rossell\'o and Maria del Mar Batle.


\begin{thebibliography}{}

\bibitem{schwinger60} J. Schwinger, Proc. Nat. Acad. Sci. U.S.A., \textbf{46}, 560, (1960)

\bibitem{ivanovic81} I. D. Ivanovi\'c, J. Phys. A, {\bf 14}, 3241, (1981)

\bibitem{wootters+89}
W. K. Wootters and B. D. Fields,
Ann. Phys. (N.Y.), {\bf 191}, 363 (1989)

\bibitem{archer05} C. Archer, J. Math. Phys. \textbf{46}, 022106 (2005)

\bibitem{planat+06} M. Planat, H. Rosu, and S. Perrine, Found. Phys. \textbf{36}, 1662 (2006)

\bibitem{cerf+02} N. Cerf, M. Bourennane, A. Karlsson, and N. Gisin: Phys.
Rev. Lett. \textbf{88}, 127902 (2002)

\bibitem{brierley09} S. Brierley: \textit{Quantum key distribution highly sensitive to eavesdropping}, arXiv:0910.2578

\bibitem{aharonov+01} Y. Aharonov and B. G. Englert, Z. Naturforsch A: Phys. Sci. \textbf{56}, 16 (2001)

\bibitem{s6} S. Brierley and S. Weigert, Phys. Rev. A \textbf{78}, 042312 (2008)

\bibitem{s7} S. Brierley and S. Weigert, Phys. Rev. A \textbf{79}, 052316 (2009)

\bibitem{s8} P. Raynal, X. L$\ddot{u}$, and B.-G. Englert, Phys. Rev. A \textbf{83}, 062303 (2011)

\bibitem{s9} D. McNulty and S. Weigert, J. Phys. A: Math. Theor. \textbf{45}, 102001 (2012)

\bibitem{complex}  W. Tadej and K. Zyczkowski, Open Sys. \& Information Dyn. \textbf{13}, 133 (2006)

\bibitem{kirkpatrick83}
S. Kirkpatrick, C. D. Gelatt Jr and M. P. Vecchi, Science {\bf 220} (4598), 671 (1983)

\bibitem{Haar33} A. Haar, Ann. Math. {\bf 34}, 147 (1933).

\bibitem{Conway90} J. Conway, {\it Course in Functional Analysis} 
(Springer-Verlag, New York, 1990)

\bibitem{Haarsimetria} M. Chaichian and R. Hagedorn, {\it Symmetries in quantum mechanics}, 
(Inst. of Phys. Publ., Bristol)

\bibitem{Mehta90} M. L. Mehta, {\it Random Matrices} (Academic, New York, 1990)

\bibitem{Hur1887} A. Hurwitz, Nachr. Ges. Wiss. G\"ott. Math.-Phys. Kl. {\bf 71} (1887)

\bibitem{Girko90} V. L. Girko, {\it Theory of Random Determinants} (Kluwer, Dordrecht, 1990)

\bibitem{metrics} T. Durt et al., Int. J. Quantum Information \textbf{8}, 535 (2010)

\bibitem{Dardo} D. Goyeneche, J. Phys. A: Math. Theor. \textbf{46}, 105301 (2013)


\bibitem{Jaming} P. Jaming et al,  J. Phys. A: Math. Theor. \textbf{42}, 245305 (2009)

\bibitem{future} J. Batle et al. Under preparation (2014)


\end{thebibliography}
\end{document}